# ANALYSIS OF POST-WET-CHEMISTRY HEAT TREATMENT EFFECTS ON NB SRF SURFACE RESISTANCE[*]

Pashupati Dhakal[#], Gianluigi Ciovati, Peter Kneisel and Ganapati Rao Myneni
Jefferson Lab, Newport News, VA 23606

*Abstract*

Most of the current research in superconducting radio frequency (SRF) cavities is focused on ways to reduce the construction and operating cost of SRF-based accelerators as well as on the development of new or improved cavity processing techniques. The increase in quality factors is the result of the reduction of the surface resistance of the materials. A recent test [1] on a 1.5 GHz single cell cavity made from ingot niobium of medium purity and heat treated at 1400 °C in a ultra-high vacuum induction furnace resulted in a residual resistance of ~ 1nΩ and a quality factor at 2.0 K increasing with field up to ~ 5×10$^{10}$ at a peak magnetic field of 90 mT. In this contribution, we present some results on the investigation of the origin of the extended $Q_0$-increase, obtained by multiple HF rinses, oxypolishing and heat treatment of "all Nb" cavities.

## INTRODUCTION

The overall performance of SRF cavities is measured by its quality factor, $Q_0 = G/R_s$, where $G$ is the geometric factor which depends on the cavity geometry and $R_s$ is the surface resistance, as a function of accelerating gradient, $E_{acc}$. Higher $Q_0$ for the reduction of cryogenic loss and higher $E_{acc}$ for the use of high energy accelerators are desired. For the continuous wave (CW) applications, the affordable cryogenic refrigeration currently limits the optimal gradient to ~20 MV/m. Thus the increase in quality factor in this gradient range is important for the efficient operation of future CW accelerators.

The surface resistance in superconducting materials is the sum of the temperature independent residual resistance and temperature dependent Bardeen-Cooper-Schrieffer (BCS) resistance. The sources of the $R_{res}$ are the trapped magnetic flux during the cavity cool down, impurities, hydrides and oxides, imperfections, and surface contamination. The BCS surface resistance results from the interaction between the RF electric field within the penetration depth and thermally activated electrons in a superconductor. The BCS surface resistance is calculated using the BCS theory and depends on superconducting material parameters. An approximated expression for the temperature range $T < T_c/2$ is given by

$$R_{BCS}(T,f) = (Af^2/T) e^{-\Delta/k_B T} \quad (1)$$

where $f$ is the resonant frequency, $\Delta$ is the superconducting gap at 0 K, $K_B$ is the Boltzmann constant and $A$ is a factor which depends on material parameters such as the coherence length ($\xi$), the London penetration depth ($\lambda_L$) and the electrons mean free path ($l$). The material parameters may vary strongly due the presence of the metallic impurities as well as defects on the surface within the RF penetration depth. Changes in these parameters affect the surface resistance of the SRF cavities and hence the quality factor. Furthermore, defects, dislocations and internal stresses (electron scatterers) also affect the superconducting properties such as the transition temperature and field of first flux penetration [2]. Dislocation sites provide pinning centers for the magnetic fluxoids, which create hot spots in SRF cavities [3]. Studies also showed that dislocations are possible nucleation sites for "etch pits", which also cause increased RF losses [4].

The BCS resistance given by Eq. (1) describe the linear BCS resistance calculated for $H_{RF} << H_c$. In case of SRF cavities where the RF magnetic field is significant ($H_{RF} \leq H_c$), the non-linearity in BCS resistance kicks in. In the presence of the RF field the coherent motion of Cooper pairs constituting the shielding current reduces the energy gap and increases the BCS loss in SRF cavities [5,6].

The increase in quality factor can be achieved by minimizing the surface resistance; both residual and BCS resistance. The current state of art cavity fabrication processes includes the bulk surface removal by buffer chemical processing (BCP) or CBP (centrifugal barrel polishing (CBP) followed by the heat treatments (600-1000 °C) to remove the gross hydrogen. A surface removal of ~20 µm either by BCP or electropolising (EP) is carried out after the heat treatment to remove the surface contaminants introduced during the heat treatment but this step can introduce the hydrogen in the cavity [7]. Earlier reports on single cell cavities showed the improvement of the quality factor without the wet chemistry after the heat treatments [1,8,9,10]. Current research and development of the SRF cavities is focused not only on achieving the higher gradient and higher quality factor but also looking to simplify the fabrication process and reducing the toxic acid during the chemistry process in bulk Nb SRF cavities. One of the processes that put forward is the bulk material removal by CBP followed by the high temperature heat treatment and high pressure rinse.

To extend the investigation on the effect of the high temperature heat treatment on the SRF cavities, a dedicated induction furnace [11] is installed at Jefferson

---

Lab. In this new furnace the heat treatment on cavity was done in two steps. Firstly, the cavity is heat treated in high vacuum at target temperature and at the end of heat treatment the furnace is purge with ultra-pure argon gas with partial pressure of ~$10^{-5}$ torr. At this temperature the only detectable gas coming from the furnace/cavity is hydrogen. The partial pressure of hydrogen is typically found to be ~$10^{-5}$-$10^{-6}$ torr. The purging of argon may prevent the reabsorption of hydrogen which is believed to one of the source of rf losses and high field non-linearities in SRF cavities. Secondly, once the cavity was cooled down to room temperature, the furnace is vented with high purity oxygen to seal the cavity surface with a protective oxygen layer against hydrogen pick-up [12,13]. As discussed in ref. [14] such a "dry" oxide layers will be beneficial with less defects than a "wet" oxide.

In this contribution, we present the summary of the recent effort in minimizing the surface resistance/rf losses in SRF cavities. Several single cell cavities were heat treated at the temperature range of 800-1600 °C and RF measurement were conducted. The cavities were chemically etched by buffer chemical polishing before the heat treatment. The temperature dependence surface resistance as well as the $Q_0$ vs $B_p$ were measured at difference temperatures.

## CAVITY TEST RESULTS

In a first set of experiments, a 1.5 GHz single cell cavity made of large grain niobium with RRR~ 200 was heat treated in the temperature range of 800-1400 °C [1]. Before each heat treatment the cavity was etched by BCP to re-establish a baseline. After each heat treatment, the cavity was cooled down in the presence of high purity Ar gas of partial pressure ~$10^{-5}$ torr and vented the cavity with high purity oxygen and soaked for ~1 hour. The cavity was then degreased in ultrasonic tank for 30 min and rinsed with high pressure deionised water. The high temperature heat treatment resulted in an increase of quality factor over the baseline for any of the heat treatment temperatures with the highest increase being observed after 1400 °C. The $R_s$ vs $T$ curve at ~10 mT were also measured in the temperature range 4.3-1.6 K during all rf test and material parameters are extracted from the fit as tabulated in Table 1. Samples analysis with surface analytical instrumentation hinted at the possibility of the presence of a small concentration (~1 at.%) of Ti within the top ~1μm layer of the surface as being involved in the large $Q_0$-improvement. Titanium evaporated from the Ti45Nb flanges of the cavity. To explore the role of the Ti contamination on the cavity performance, two approaches were followed:

- The cavity was subjected to multiple steps of nanoremoval from the surface by rinsing with HF and oxypolishing [15] with cryogenic rf tests in between. The summary of results are presented in Fig. 1. It is shown that the extended Q-rise is still present even after the removal of ~100nm from the inner surface of the cavity. The cavity was then subjected to ~30 μm EP surface polishing and the extended $Q$-rise was eleminated while reaching the cavity ~25 MV/m.

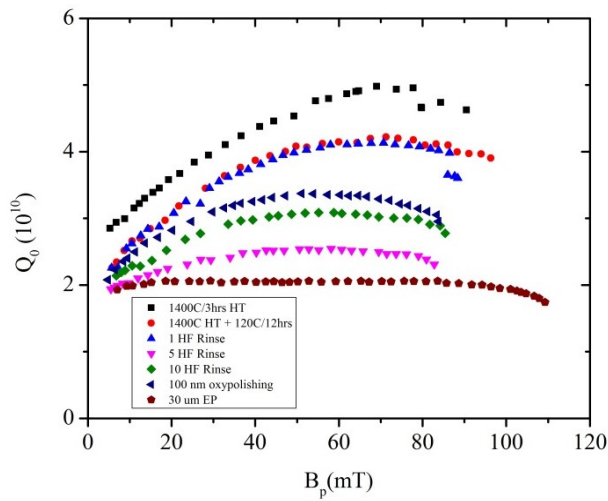

Figure. 1. The $Q_0$ vs $B_p$ at 2K for the cavity with Ti45Nb that subjected to high temperature heat treatment and subjected to multiple HF rinsing, oxypolishing and EP The ratio $B_p/E_{acc}$ is 4.43 mT/(MV/m) for this cavity shape.

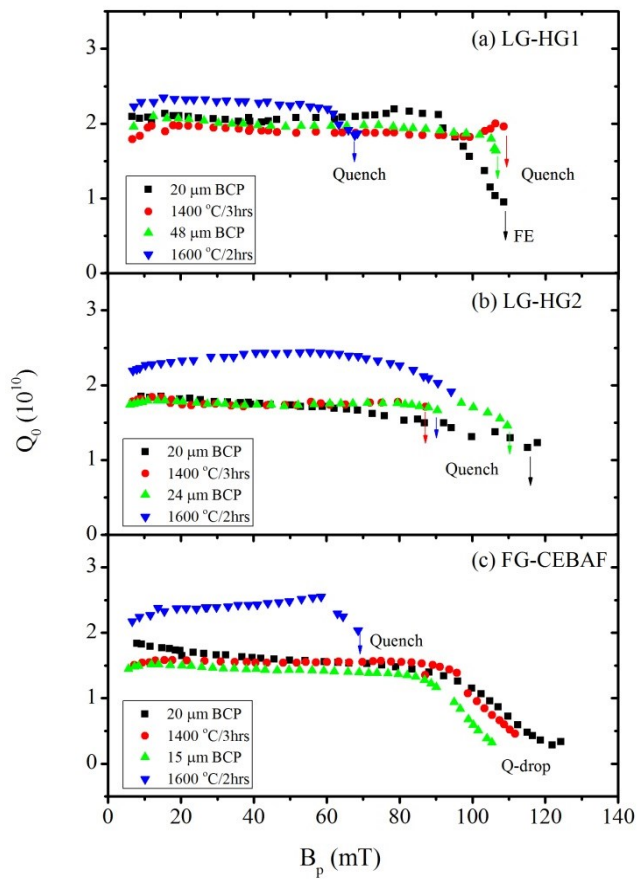

Figure 2: The $Q_0$ vs $B_p$ at 2K for "all Nb" cavities subjected to high temperature heat treatment.

- Two large-grain (RRR~200) and one fine-grain (RRR>300) single-cell cavities with pure Nb flanges were heat-treated at 1400 °C and 1600 °C after etching with BCP 1:1:2. Although the $Q_0$ at 2.0K, 90 mT improvement in $Q_0$, no extended $Q_0$-rise as that shown in Fig 1 was observed in any of these cavities. The rf test results at 2.0 K for these "all Nb" cavities are shown in Fig. 2.

Table 1. The superconducting parameters extracted from the fitting of $R_s$ vs T measurements form 1.5 GHz single cell cavities namely: large grain CEBAF shape (LG-CEBAF, $B_p/E_{acc}$ = 4.43 mT/(MV/m)), large grain high gradient shape (LG-HG, $B_p/E_{acc}$ = 4.469 mT/(MV/m)), fine grain (FG) and reactor grade (RG). The cavity LG-CEBAF has Ti45Nb flanges.

| Cavity Shape | Treatment | $R_s$ (n$\Omega$) | $\Delta/K_BT_c$ | $l$ (nm) | $B_{p,max}$ (mT) | $Q_0$ (2K, 90mT) |
|---|---|---|---|---|---|---|
| LG-CEBAF | 20 μm BCP | 2.0 ± 0.3 | 1.87 ± 0.02 | 303 ± 85 | 100 ± 6 | $(2.0 \pm 0.3) \times 10^{10}$ |
| | 20 μm BCP + 1400C/3hrs | 1.0 ± 0.2 | 1.90 ± 0.01 | 76 ± 17 | 91 ± 9 | $(5.0 \pm 1.0) \times 10^{10}$ |
| | 20 μm BCP + 1400C/3hrs +120C/12hrs | 2.8 ± 0.2 | 1.93 ± 0.02 | 7 ± 1 | 96 ± 7 | $(4.0 \pm 0.7) \times 10^{10}$ |
| | +1 HF Rinse | 1.4 ± 0.2 | 1.86 ± 0.01 | 29 ± 47 | 88 ± 7 | $(3.6 \pm 0.6) \times 10^{10}$* |
| | +5 HF Rinse | 5.1 ± 0.1 | 1.90 ± 0.01 | 84 ± 14 | 83 ± 5 | $(2.3 \pm 0.3) \times 10^{10}$* |
| | +10 HF Rinse | 2.8 ± 0.1 | 1.88 ± 0.02 | 57 ± 23 | 85 ± 5 | $(2.8 \pm 0.3) \times 10^{10}$* |
| | +100 nm oxypolishing | 2.3 ± 0.2 | 1.87 ± 0.02 | 52 ± 25 | 84 ± 4 | $(2.8 \pm 0.3) \times 10^{10}$* |
| | + 30μm EP | 3.5 ± 0.3 | 1.80 ± 0.02 | 81 ± 26 | 110 ± 6 | $(2.0 \pm 0.2) \times 10^{10}$ |
| LG-HG1 | 20 μm BCP | 0.05 ± 0.75 | 1.78 ± 0.02 | 44 ± 24 | 124 ± 9 | $(1.5 \pm 0.2) \times 10^{10}$ |
| | 20 μm BCP + 1400C/3hrs | 4.6 ± 0.2 | 1.87 ± 0.01 | 58 ± 12 | 92 ± 5 | $(1.7 \pm 0.2) \times 10^{10}$ |
| | 24 μm BCP | 2.7 ± 0.2 | 1.83 ± 0.03 | 77 ± 45 | 115 ± 5 | $(1.7 \pm 0.2) \times 10^{10}$ |
| | 24 μm BCP + 1600C/2hrs | 2.7 ± 0.1 | 1.88 ± 0.02 | 55 ± 29 | 98 ± 4 | $(2.1 \pm 0.2) \times 10^{10}$ |
| LG-HG2 | 20 μm BCP | 0.0 ± 0.6 | 1.79 ± 0.01 | 68 ± 20 | 114 ± 11 | $(2.1 \pm 0.4) \times 10^{10}$ |
| | 20 μm BCP + 1400C/3hrs | 2.6 ± 0.2 | 1.87 ± 0.01 | 190 ± 30 | 120 ± 6 | $(1.9 \pm 0.2) \times 10^{10}$ |
| | 48 μm BCP | 2.9 ± 0.9 | 1.91 ± 0.02 | 551 ± 201 | 112 ± 6 | $(1.9 \pm 0.2) \times 10^{10}$ |
| | 48 μm BCP + 1600C/2hrs | 1.6 ± 0.2 | 1.82 ± 0.02 | 62 ± 18 | 72 ± 4 | $(1.9 \pm 0.2) \times 10^{10}$* |
| FG-CEBAF | 20 μm BCP | n/a | n/a | n/a | 124 ± 9 | $(1.4 \pm 0.2) \times 10^{10}$ |
| | 20 μm BCP + 1400C/3hrs | 2.9 ± 0.2 | 1.83 ± 0.01 | 354 ± 65 | 112 ± 7 | $(1.5 \pm 0.2) \times 10^{10}$ |
| | 15 μm BCP | 0.2 ± 0.4 | 1.75 ± 0.02 | 124 ± 41 | 105 ± 6 | $(1.2 \pm 0.1) \times 10^{10}$ |
| | 15 μm BCP + 1600C/2hrs | 0.7 ± 0.2 | 1.86 ± 0.02 | 203 ± 61 | 70 ± 4 | $(2.3 \pm 0.3) \times 10^{10}$* |
| RG-CEBAF | 20 μm BCP | n/a | n/a | n/a | 106 ± 5 | $(1.0 \pm 0.1) \times 10^{10}$ |
| | 20 μm BCP + 1400C/3hrs | 4.0 ± 0.6 | 1.87 ± 0.02 | 84 ± 27 | 89 ± 7 | $(2.1 \pm 0.4) \times 10^{10}$* |

*at highest field

In addition one single cell fabricated from reactor grade niobium (RRR~40) was heat treated at 1400 °C for 3 hours. The result of the rf test at 2K is shown in Fig. 3. The cavity reached ~20 MV/m with $Q_0 = (2.1 \pm 0.4) \times 10^{10}$. The cavity was post-purified in Ti box at 1250 °C for 3 hours and ~30 μm surface was removed by BCP before the baseline test. Interestingly, the $Q$-rise was observed after 1400 °C heat treatment but only up to ~35 mT.

## DISCUSSIONS

The increase in quality factor was observed for the cavity heat treated at very high temperature. This increase in the quality factor is basically due to the reduction of the residual resistance and BCS resistance in some cases. A possible cause for $Q$-slopes both at high and medium field could be the precipitation of niobium hydride islands [16] within the rf penetration depth and $Q$-disease occurs if the hydrogen concentration in Nb is greater than 10 wt. ppm. To minimize if not eliminate hydride formation; the mobile hydrogen concentration has to be kept as low as possible. This can be done either by complete degassing or providing the trapping sites for hydrogen within the Nb lattice to reduce hydrogen mobility. The heat treatments at 800-1200 °C reduce the H concentration and improved $Q_0$-values were obtained both in these and earlier studies [1,9,10]. The presence of Ti and O at the Nb surface after the 1400 °C could play an important role as they provide very effective trapping centers for hydrogen. The reduction of H content combined with the introduction of H-trapping sites, as obtained by the 1400 °C heat treatment causes the reduction in surface resistance. Therefore, the presence of trapping centers for hydrogen within the rf penetration depth might be beneficial in achieving higher $Q_0$ values in SRF cavities.

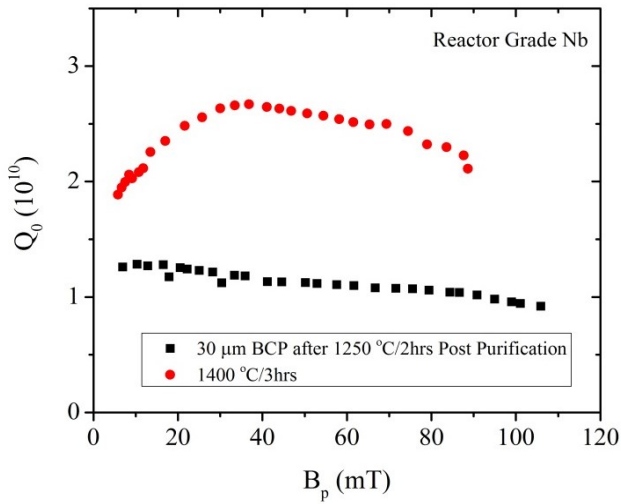

Figure 3: The $Q_0$ vs $B_p$ at 2K for reactor grade cavity (RG-CEBAF) subjected to 1400 °C heat treatment. The ratio $B_p/E_{acc}$ is 4.43 mT/(MV/m) for this cavity shape.

Another outstanding feature of these rf tests is the extended $Q_0$-rise occurring only for the cavity heat treated at 1400 °C with Ti45Nb flanges. An explanation of the low field $Q$-rise (decrease in surface resistance) was proposed by J. Halbritter [17]. According to this model, an additional $NbO_x$ clusters grown at the surface are not in thermal equilibrium with the surrounding niobium causing the decrease in surface resistance. However, he predicted that the equilibrium will be achieved ~12 mT at which the surface resistance starts to increase. Some earlier experimental data fits well [18] according the model described by Halbritter. Here the low field $Q_0$-rise extended up to ~60 mT for the cavity heat treated at 1400 °C and similar rise in $Q_0$ was observed at Fermi Lab where the cavities were heat treated at 1000 °C in the presence of higher partial pressure of nitrogen (~$10^{-2}$ torr) [19]. However, the cavities were subjected to EP surface treatment after the high temperature heat treatment contrary to the current study where no post-chemistry were performed after the heat treatment. In past, the extended rise in $Q_0$ was observed in 3 GHz cavities which were oxidized at high temperature [20] may support the Halbritter's idea of $NbO_x$ clusters at the surface of SRF cavities. However, the rise in $Q_0$-values in cavities doped with Ti as well as nitrogen treated cavities may require further detail investigation.

As discussed earlier several factors play a significant role in the surface resistance of the SRF cavities. The surface oxides, hydrides, surface morphologies, grain boundaries, etch pits and so on. No theoretical models exist that correlates all these contributing factors in the presence of the RF field. Earlier calculation on the surface resistance of thin film superconductor taking in to account of the quasi-particle redistribution showed the decrease in surface resistance with the increase in microwave power level [21]. The contribution to the microwave surface resistance is much higher due to the quasi-particle redistribution than the enhancement of the energy gap. Interestingly, recent numerical calculation based on the Mattis and Bardeen theory modified to account for moving Cooper pairs under the action of the rf field, describe the RF field dependence of quality factor of 1.5 GHz SRF cavity quite well [22].

The same heat treatment procedure was applied to the cavities fabricated with "all Nb" flanges. The cavities were fabricated with large grain, high RRR fine grain and low RRR reactor grade. The improvement in $Q_0$ up to 70% in medium field was observed in all cases. The extended $Q$-rise is also observed in the cavity fabricated from reactor grade niobium. It should be noted that the cavity was purified using the Ti at temperature 1250 °C and BCP ~30 μm was done both inside and outside of the cavity prior to the baseline test.

## SUMMARY AND FUTURE WORKS

The high temperature heat treatment of the SRF cavities with no post-chemistry resulted in the overall increase in quality factor at 2.0 K and accelerating gradient of ~20 MV/m. The elimination of the surface material removal (after EP or BCP) not only eliminates an extra processing step but also minimize the risk of having hydrogen re-absorption. Further studies will focus on:

- The reliability of the proposed process to otain cavities with $Q_0$(2.0K, 90 mT, 1.5 GHz) of ~ $4 \times 10^{10}$.
- Further investigation of the impact of surface impurities on the $Q_0(B_p)$ curve through cavity rf tests and studies on small samples cutout from the cavities.

## ACKNOWLEDGEMENT

We would like to acknowledge JLAB SRF staffs for the technical and cryogenic supports.